\documentclass[conference]{IEEEtran}
\IEEEoverridecommandlockouts
\usepackage{cite}
\usepackage{amsmath,amssymb,amsfonts}
\usepackage{algorithm}
\usepackage{algorithmic}
\usepackage{amsthm}

\usepackage{graphicx}
\usepackage{textcomp}
\usepackage{xcolor}

\def\BibTeX{{\rm B\kern-.05em{\sc i\kern-.025em b}\kern-.08em
    T\kern-.1667em\lower.7ex\hbox{E}\kern-.125emX}}
\begin{document}

\title{ProHD: Projection-Based Hausdorff Distance Approximation}

\author{
    \IEEEauthorblockN{Jiuzhou Fu, Luanzheng Guo, Nathan R. Tallent, Dongfang Zhao\thanks{Corresponding author: dzhao@uw.edu}}
    \IEEEauthorblockA{University of Washington\\
    Email: jiuzhou@uw.edu, dzhao@uw.edu}
}

\author{%
  \IEEEauthorblockN{Jiuzhou Fu$^{1}$, Luanzheng Guo$^2$, Nathan R. Tallent$^2$, Dongfang Zhao$^1$\thanks{Corresponding author: dzhao@uw.edu}}
  \IEEEauthorblockA{$^1$University of Washington, $^2$Pacific Northwest National Laboratory\\
  Emails: \{jiuzhou, dzhao\}@uw.edu; 
  \{lenny.guo, tallent\}@pnnl.gov
  }
}

\maketitle 

\begin{abstract}
The Hausdorff distance (HD) is a robust measure of set dissimilarity, but computing it exactly on large, high-dimensional datasets is prohibitively expensive. We propose \textbf{ProHD}, a projection-guided approximation algorithm that dramatically accelerates HD computation while maintaining high accuracy. ProHD identifies a small subset of candidate “extreme” points by projecting the data onto a few informative directions (such as the centroid axis and top principal components) and computing the HD on this subset. This approach guarantees an underestimate of the true HD with a bounded additive error and typically achieves results within a few percent of the exact value. In extensive experiments on image, physics, and synthetic datasets (up to two million points in $D=256$), ProHD runs 10–100$\times$ faster than exact algorithms while attaining 5–20$\times$ lower error than random sampling-based approximations. Our method enables practical HD calculations in scenarios like large vector databases and streaming data, where quick and reliable set distance estimation is needed.
\end{abstract}


\section{Introduction}\label{sec:introduction}

\subsection{Background and Motivation}
The Hausdorff distance (HD) is a fundamental geometric measure extensively utilized to quantify the dissimilarity between point clouds, shapes, and data embeddings across diverse fields such as computer vision\cite{visionrucklidge1996efficient,visionzhao2005new}, graphics\cite{visionDIRAC1953888,visionFEYNMAN1963118}, computational biology and medicine\cite{visionMRI10.1007/978-3-540-85988-8_49,visionMRI6975210}, and machine learning\cite{machinelearningaksoy2019relative,machinelearningpiramuthu1999hausdorff}. Its robustness against non-uniform sampling and missing correspondences makes it a particularly compelling choice for critical tasks including shape registration, LiDAR scan alignment, 3D model verification, trajectory analysis, and evaluating robustness of machine learning models against adversarial perturbations.

Despite its widespread use, the standard computational method for HD has a significant practical limitation: its complexity is quadratic with respect to the number of points ($n$) and linear with respect to dimensionality ($D$). This makes exact computation computationally infeasible for modern, large-scale, high-dimensional datasets frequently encountered in real-world scenarios. For example, tasks involving hundreds of thousands or millions of points quickly become computationally prohibitive, requiring substantial processing time even with optimized exact methods like KD-tree-based indexing, pruning strategies, or branch-and-bound algorithms.

Consequently, approximating the HD quickly and relatively accurately has emerged as a critical area of research. Common approaches have traditionally relied on uniform or stratified random sampling and global dimensionality reduction techniques such as Principal Component Analysis (PCA)\cite{pca,pmlr-v97-simonov19a,marukatat2023tutorial} or random projections\cite{bingham2001random,vempala2005random}. While sampling methods are computationally inexpensive, they typically yield highly variable errors due to their inherent randomness and inability to focus on geometric extremes that define the true Hausdorff distance. Global dimensionality reduction methods preserve distances in expectation but often compress the extremes that crucially determine HD, leading to inaccuracies especially in moderate to high-dimensional spaces.

Therefore, there is a strong motivation to develop methods that can reliably and rapidly estimate the Hausdorff distance with significantly reduced computational overhead compared to exact algorithms, while simultaneously offering substantially improved accuracy compared to traditional approximate methods. Additionally, the rise of large-scale vector databases and embedding repositories has created demand for fast metrics between sets of vectors. A quick Hausdorff distance approximation can, for instance, help compare groups of high-dimensional embeddings or track distributional drift in a vector database, supporting data analysis and anomaly detection at scale.

\subsection{Proposed Work}
To address these limitations, we introduce \textbf{ProHD}, a projection-guided Hausdorff distance approximation method specifically designed for large-scale, high-dimensional datasets. Our approach leverages a novel selection strategy rooted in geometric insights about the distribution of extreme points that typically govern the Hausdorff distance.

The core idea behind \textbf{ProHD} is based on two key observations: (1) the points contributing to the Hausdorff distance typically lie near the extremes of the combined point clouds along certain discriminative directions, and (2) these directions can be effectively identified through projection methods, specifically centroid-based and PCA-based projections.

Specifically, \textbf{ProHD} projects the union of the two point clouds onto the centroid direction (connecting the centroids of the two clouds) and onto the top principal components derived from the combined dataset. From each direction, it selects points at the extremes, retaining only the top- and bottom-\hspace{0.2em}$\alpha$-fraction along each projected axis. The resulting subsets thus contain the points most likely to determine the actual Hausdorff distance.

These significantly smaller subsets are then used to compute an exact Hausdorff distance using a fast ANN-based indexing method, such as the Faiss flat index. Because the subset sizes scale favorably (roughly ), our method achieves a substantial reduction in computational time compared to exact approaches, while providing a demonstrably higher accuracy compared to random sampling or systematic sampling methods.

\subsection{Contributions}
The primary contributions of our work, \textbf{ProHD}, include:

\begin{enumerate}
\item \textbf{Fast and Accurate Approximation:} \textbf{ProHD} significantly accelerates Hausdorff distance estimation, offering runtimes that are typically – faster than exact methods like KD-trees, Z-order curves, and pruning-based algorithms. Simultaneously, it achieves a relative accuracy substantially better than traditional approximate methods such as random and systematic sampling.

\item \textbf{Theoretical Guarantees:} We provide rigorous theoretical bounds on the accuracy of our method, showing that our approach always underestimates the true Hausdorff distance by at most a deterministic additive error of \(2\min_u\delta(u)\). Additionally, we demonstrate a monotonic convergence towards the true Hausdorff distance as additional directions are included, establishing a strong theoretical underpinning for the method’s reliability.

\item \textbf{Scalability through Extensive Evaluation:} Our comprehensive empirical evaluation demonstrates the method’s scalability and effectiveness across various datasets, including image data (MNIST and CIFAR-10), physics datasets (Higgs boson data), and synthetic point clouds with up to two million points and dimensions up to \(D=256\). These evaluations confirm that \textbf{ProHD} not only scales gracefully with dataset size and dimensionality but also consistently delivers accuracy significantly superior to standard approximate approaches.

\end{enumerate}

Through these contributions, \textbf{ProHD} significantly expands the feasibility of using Hausdorff distance computations in large-scale applications, balancing computational efficiency with accuracy, and opening new possibilities for scalable, precise geometric analyses.

\section*{Related Work}
\label{sec:related work}
The \emph{Hausdorff distance} (HD) is one of the oldest set distances in mathematics and remains the gold standard for quantifying the maximal mismatch between two shapes. Unfortunately the naïve algorithm is quadratic in the number of points, $O(N^2D)$, as first analysed in the early computational-geometry literature \cite{naiveH10.1007/0-387-33006-2_4}.  
Research over the last four decades can be grouped into three broad directions:  
(i) exact but faster-than-quadratic methods,  
(ii) controlled approximations that trade a guaranteed $\varepsilon$ error for sub-quadratic time, and  
(iii) GPU-accelerated or domain-specific implementations that exploit hardware or application structure.  
Below we review each strand and point to more than thirty representative papers.

\paragraph{Exact Algorithms.}
\emph{Brute force and early pruning.}  
Brute-force scanning is still widely used as a baseline; modern SIMD implementations reduce the constant factors but not the asymptotic $O(|X||Y|)$ cost. Several pruning strategies have been proposed to skip large subsets of the search space. Early-break Hausdorff (\emph{EBHD}) uses randomised ordering plus an \emph{early break} rule to cut off the inner loop once the current best bound is violated \cite{EBHD1ChauDang-Nguyen2020,EBHD27053955}. Local-Start Search (\emph{LSS}) records the previous break position and restarts the next query in its neighbourhood, obtaining large savings on partially overlapping sets \cite{10.1016/j.patcog.2017.02.013}. Diffusion-Search HD combines a Z-order curve for sparse data with an octree walk for dense data \cite{ZHD8125673}. The recent \emph{PRSR} (points-ruling-out and systematic random sampling) algorithm focuses on pruning the \emph{outer} loop and shows further gains on million-point clouds \cite{RYU2021107857}.  

\emph{Tree-based spatial indexes.}  
Space-partitioning trees accelerate the nearest-neighbour sub-problem that lies at the core of HD. Classical k-d trees \cite{EBHD310.1145/361002.361007} and their statistical variants \cite{TreeJames2023} work well up to medium dimension, but suffer from the curse of dimensionality \cite{Tree6065061}. R-trees and their aggregated-nearest-neighbour extension were the first to support exact HD queries in spatial databases \cite{Papadias2005}. CD-Tree \cite{EBHD4doi:10.3233/IDA-150418} integrates clustering with dynamic insertions, whereas octree-based NOHD/OHD prune 3-D volumes via bounding boxes \cite{10.3233/ICA-170544}.  

\emph{Exact ANN back-ends.}  
A different line of work treats HD as \emph{many} nearest-neighbour calls and simply swaps in a high-performance library. With FAISS’ flat (brute-force but SIMD) index configured for $k{=}1$, multi-threaded CPU implementations reach near-linear scaling in practice while remaining exact \cite{Johnson2017}. Multi-vector databases expose the same idea at data-platform scale \cite{ANNunknown}.  

\paragraph{Approximate Hausdorff Distance.}

\emph{Sampling and randomised pruning.}  
Uniform or stratified sampling of each cloud is still used in graphics pipelines, but its worst-case error is uncontrolled. Systematic random sampling combined with PRSR’s ruling-out mechanism offers an empirically tighter bound \cite{RYU2021107857}.  

\emph{$(1{+}\varepsilon)$-approximation schemes.}  
Indyk and Motwani recently showed how to approximate HD between patterns under rigid motions via locality-sensitive hashing \cite{Indyk1999}. For low-doubling-dimension metrics, Chubet et.al give linear-time preprocessing and $(1{+}\varepsilon)$ queries by applying dual-tree \emph{greedy} nets\cite{chubet2025approximatingdirectedhausdorffdistance} to obtain linear query time and supports $k$-partial HD (outlier trimming) with no extra cost. Yet no algorithm implementation was given in the paper.  

\emph{Budgeted and partial distances.}  
Huttenlocher’s \emph{partial HD} introduced the idea of ignoring $k$ worst outliers for robustness \cite{Huttenlocher1993}. Har-Peled \textit{et al.} \cite{HarPeled2023} formalise the trade-off between computation budget and approximation error, providing tight complexity bounds.  

\paragraph{GPU-Accelerated and Domain-Specific Variants.}

\emph{Surface-oriented GPU pipelines.}  
For free-form geometry where the input is a pair of trimmed NURBS patches, Krishnamurthy \emph{et al.} developed a two-stage GPU pipeline that first builds bounding-box hierarchies entirely on the device and then performs a breadth-first culling traversal before refining the surviving leaf pairs at full resolution \cite{Yang2011GPU}. Follow-up work showed that the same idea can be pushed further with an iterative GPU kernel that numerically converges on the maximal patch-to-patch separation, reducing CPU $\Leftrightarrow$ GPU synchronisation overhead \cite{HANNIEL2012255}. These approaches achieve interactive rates for moderately tessellated models, but their error bounds grow with surface curvature and the memory footprint scales poorly when many small Bézier sub-patches are required, making them less attractive for highly detailed CAD data.  

\emph{Interactive polygonal models.}  
On triangle meshes, Tang and colleagues used a GPU-friendly bounding-volume hierarchy (BVH) combined with cross-culling of triangle pairs to deliver frame-rate Hausdorff queries inside virtual-reality applications \cite{Tang2009}. While extremely fast, the method is approximate: the BVH tolerance must be tuned and becomes scene dependent, and the implementation is restricted to closed, manifold surfaces.  

\emph{Diffusion-search for large point sets.}  
The diffusion-search framework of Zhang \emph{et al.} tackles point-cloud instances by starting each inner nearest-neighbour search at the “break-point’’ of the previous iteration, greatly reducing average work. Two specialised kernels are provided: ZHD for sparse clouds ordered along a Morton (Z-order) curve and OHD for dense voxel data indexed in an octree \cite{ZHD8125673}. Although the authors present a CPU implementation, the preprocessing stages—Morton-code generation, radix sort and octree construction—are most efficient on GPUs, and the paper explicitly relies on CUDA primitives such as Thrust for practical throughput \cite{ZHD8125673}. Consequently, the algorithm inherits the usual GPU constraints (device memory limits and expensive host–device transfers) and its performance degrades when the point distribution violates Morton-locality assumptions.

\section{Methodology}
\label{sec:method}
This section introduces our proposed \textbf{ProHD} approach. We begin with background concepts and notation for Hausdorff distance approximation, then describe the intuition behind our algorithm and outline the key steps. We also analyze its computational complexity and provide theoretical guarantees on approximation accuracy.

\subsection{Background and Notation}
Our objective is to approximate the undirected Hausdorff distance
\begin{align}
    \label{eq:H_def}
  &H(A,B) \;=\; \max\bigl\{\,h(A,B),\,h(B,A)\bigr\}, \\
  &\text{where}\quad h(A,B) \;=\; \max_{a\in A}\min_{b\in B}\|\,a - b\|_2,
\end{align}
for two point clouds \(A,B\subset\mathbb{R}^D\), in a fraction of the time required by exact algorithms. Rather than inspecting every point–pair, we identify a small “extreme” subset of \(A\) and \(B\) by projecting onto a handful of carefully chosen directions: the line through their centroids plus the top principal components of \(A\cup B\). Extracting the points whose projections lie in the top and bottom \(\alpha\)-fractions along each direction yields two tiny sets \(A_{\mathrm{sel}}\subset A\), \(B_{\mathrm{sel}}\subset B\). We then compute 
\[
  \widehat H(A,B)
  \;=\;
  \max\Bigl\{\,\max_{a\in A_{\mathrm{sel}}}\min_{b\in B_{\mathrm{sel}}}\|a-b\|_2,\;
               \max_{b\in B_{\mathrm{sel}}}\min_{a\in A_{\mathrm{sel}}}\|b-a\|_2\Bigr\}
\]
via a fast ANN routine (e.g., Faiss). Because the selected directions “cover” the outer hull of each cloud, \(\widehat H(A,B)\) underestimates \(H(A,B)\) by at most an additive term that depends on how far any point lies from its projection onto the best direction. As we show below, including both the centroid direction and a few principal-component directions guarantees a low approximation error even in high dimensions.

Let
\[
  A = \{\,a_1,\dots,a_{n_A}\}, 
  \quad 
  B = \{\,b_1,\dots,b_{n_B}\}, 
  \quad 
  a_i,b_j\in\mathbb{R}^D,
\]
and denote by \(\|\cdot\|\) the Euclidean norm. The directed Hausdorff distance \(h(A,B)\) and undirected Hausdorff distance \(H(A,B)\) are defined in \eqref{eq:H_def}. We write \(\widehat{h}(A,B)\) and \(\widehat{H}(A,B)\) for the directed and undirected Hausdorff distances computed on a small subset of \(A\) and \(B\). Given a unit vector \(u\in\mathbb{R}^D\), let 
\[
  \pi_u(p) \;=\; p^\top u
  \quad\bigl(p\in A\cup B\bigr)
\]
be the projection of \(p\) onto the line spanned by \(u\). Define
\begin{equation}
  \label{eq:delta_def}
  \delta(u) 
  \;=\; 
  \max_{\,p\in A\cup B}\Bigl\|\,p - \bigl(\pi_u(p)\bigr)\,u\Bigr\|_2
  \;=\; 
  \max_{\,p\in A\cup B}\|\,\Pi_{u^\perp}(p)\|\!,
\end{equation}
the maximum distance from any point to its projection onto direction \(u\). Finally, let
\[
  m \;=\; \bigl\lfloor\sqrt{D}\bigr\rfloor,
  \quad
  \alpha \in (0,1)
\]
be parameters chosen at runtime (we typically set \(\alpha=0.01\)).

\subsection{Intuition}
Computing \(H(A,B)\) exactly requires checking the farthest neighbor of each point, which is \(O(n^2)\) in the worst case. Instead, observe that projecting the point sets onto any fixed unit vector \(u\) yields a one-dimensional Hausdorff distance $H_u(A,B)$ that is guaranteed to be no larger than the true $H(A,B)$. Intuitively, $H_u(A,B)$ ignores the component of each point’s displacement orthogonal to $u$, so it underestimates the full Hausdorff distance. The difference is bounded by how far points deviate from the direction $u$, captured by $\delta(u)$ (Equation~\ref{eq:delta_def}). In fact, we have $H(A,B) \le H_u(A,B) + 2\,\delta(u)$.

Thus, if we can identify the direction $u$ for which $\delta(u)$ is minimal, the corresponding $H_u(A,B)$ will be very close to the true $H(A,B)$. In practice, the centroid direction often captures a large separation between $A$ and $B$, but when $A\cup B$ lies in a rotated or elongated subspace, the top principal component(s) yield even smaller $\delta(u)$.

Rather than computing $H_u$ on all of $A\cup B$ (which is still expensive), we note that
\[
  H_u(A,B) \;=\; H_u\bigl(A_{\mathrm{ext}},\,B_{\mathrm{ext}}\bigr),
\]
where $A_{\mathrm{ext}}\subset A$ consists exactly of those points whose projection lies among the top-$\alpha$ or bottom-$\alpha$ coordinates along $u$. By keeping only those “extreme” points for each chosen direction, the 1D Hausdorff distances along those directions are preserved. Taking the union of these extremes for all directions yields two tiny subsets $A_{\mathrm{sel}}$ and $B_{\mathrm{sel}}$. Computing $H(A_{\mathrm{sel}},B_{\mathrm{sel}})$ (via a fast ANN) recovers 
\[
  \widehat H(A,B) 
  \;=\; \max_{u\in\{u^{(0)},\dots,u^{(m)}\}}H_u(A,B)
\]
exactly, and we still have
\begin{equation}
  \label{eq:final_bound}
  \widehat H(A,B)\;\le\;H(A,B)\;\le\;\widehat H(A,B)\;+\;2\,\min_{0\le\ell\le m}\delta\bigl(u^{(\ell)}\bigr).
\end{equation}
Below we give pseudocode for selecting these extremes and assembling the final approximation.

\subsection{Algorithms}
We define two helper routines (centroid-based selection and PCA-based selection) and then present the main procedure. In each algorithm, let \(\alpha\) denote the fraction of points kept per direction; in the main routine we set \(\alpha' = \alpha/m\) for PCA directions so that the total fraction does not exceed \(\alpha\cdot(1 + m)\).

\subsubsection{Centroid-Projection Selection}
This routine picks those points whose projections onto the centroid-direction vector lie in the top or bottom \(\alpha\)-fraction.

\begin{algorithm}[H]
  \footnotesize
  \caption{CentroidIndices\(\bigl(X,Y,\alpha\bigr)\)}
  \label{alg:centroid}
  \begin{algorithmic}[1]
    \REQUIRE \(X\in\mathbb{R}^{n_X\times D},\,Y\in\mathbb{R}^{n_Y\times D},\,\alpha\in(0,1).\)
    \ENSURE \(\mathcal{I}_X\subseteq\{1,\dots,n_X\},\;\mathcal{I}_Y\subseteq\{1,\dots,n_Y\}.\)
    \STATE Compute centroids
      \[
        \bar x = \tfrac{1}{n_X}\sum_{i=1}^{n_X}X_i,\quad
        \bar y = \tfrac{1}{n_Y}\sum_{j=1}^{n_Y}Y_j.
      \]
    \STATE Let \(u = \bar y - \bar x\). If \(\|u\|<10^{-9}\), set \(u = e_1\); otherwise normalize \(u\leftarrow u/\|u\|\).
    \FOR{\(i=1\) \TO \(n_X\)} 
      \STATE \(p_X[i] \leftarrow X_i^\top\,u\).
    \ENDFOR
    \FOR{\(j=1\) \TO \(n_Y\)} 
      \STATE \(p_Y[j]\leftarrow Y_j^\top\,u\).
    \ENDFOR
    \STATE Set \(k_X = \max\{1,\lfloor\alpha\,n_X\rfloor\},\;k_Y = \max\{1,\lfloor\alpha\,n_Y\rfloor\}.\)
    \STATE Let \(s_X = \mathrm{argsort}(p_X)\in\{1,\dots,n_X\}\) (increasing), and similarly \(s_Y=\mathrm{argsort}(p_Y).\)
    \STATE 
      \(\mathcal{I}_X 
        \leftarrow \bigl\{\,s_X[1],\dots,s_X[k_X]\bigr\}\,\cup\bigl\{\,s_X[n_X-k_X+1],\dots,s_X[n_X]\bigr\},\) 
      then keep the unique indices.
    \STATE 
      \(\mathcal{I}_Y 
        \leftarrow \bigl\{\,s_Y[1],\dots,s_Y[k_Y]\bigr\}\,\cup\bigl\{\,s_Y[n_Y-k_Y+1],\dots,s_Y[n_Y]\bigr\},\) 
      then keep the unique indices.
    \RETURN \(\mathcal{I}_X,\;\mathcal{I}_Y.\)
  \end{algorithmic}
\end{algorithm}

We compute the vector from $X$’s centroid to $Y$’s centroid, normalize it to $u$, and project all points in $X$ and $Y$ onto $u$. Sorting these scalars, we keep the smallest $k_X$ and largest $k_X$ points from $X$, and similarly for $Y$. Those indices become $\mathcal{I}_X$ and $\mathcal{I}_Y$. Intuitively, they lie near the “front” or “back” of each cloud along the centroid line.

\subsubsection{PCA-Projection Selection}
This routine finds the top $m$ principal components of all points $(X;Y)$, then for each component selects extremes in the same way as above.

\begin{algorithm}[H]
  \footnotesize
  \caption{PCAProjIndices\(\bigl(X,Y,\alpha,m\bigr)\)}
  \label{alg:pca}
  \begin{algorithmic}[1]
    \REQUIRE \(X\in\mathbb{R}^{n_X\times D},\,Y\in\mathbb{R}^{n_Y\times D},\,\alpha\in(0,1),\,m\le D.\)
    \ENSURE \(\mathcal{I}'_X\subseteq\{1,\dots,n_X\},\;\mathcal{I}'_Y\subseteq\{1,\dots,n_Y\}.\)
    \STATE Stack points $\widetilde Z = [X;Y]\in\mathbb{R}^{(n_X+n_Y)\times D}$.
    \STATE Compute the top $m$ principal components $\{u^{(1)},\dots,u^{(m)}\}$ of $\widetilde Z$.  
    \STATE Initialize $\mathcal{I}'_X = \emptyset,\;\mathcal{I}'_Y = \emptyset.$
    \FOR{$\ell = 1$ \TO $m$}
      \STATE Let $u \leftarrow u^{(\ell)}/\|u^{(\ell)}\|$.
      \FOR{$i=1$ \TO $n_X$} 
        \STATE $q_X[i]\leftarrow X_i^\top\,u$.  
      \ENDFOR
      \FOR{$j=1$ \TO $n_Y$} 
        \STATE $q_Y[j]\leftarrow Y_j^\top\,u$.  
      \ENDFOR
      \STATE $k_X = \max\{1,\lfloor\alpha\,n_X\rfloor\},\quad k_Y = \max\{1,\lfloor\alpha\,n_Y\rfloor\}$.
      \STATE $r_X = \mathrm{argsort}(q_X),\quad r_Y = \mathrm{argsort}(q_Y)$.
      \STATE 
        $\mathcal{I}'_X 
          \leftarrow \mathcal{I}'_X \cup \{r_X[1],\dots,r_X[k_X]\}\cup\{r_X[n_X-k_X+1],\dots,r_X[n_X]\}$.
      \STATE 
        $\mathcal{I}'_Y 
          \leftarrow \mathcal{I}'_Y \cup \{r_Y[1],\dots,r_Y[k_Y]\}\cup\{r_Y[n_Y-k_Y+1],\dots,r_Y[n_Y]\}$.
    \ENDFOR
    \STATE Replace $\mathcal{I}'_X,\mathcal{I}'_Y$ by their unique elements.
    \RETURN $\mathcal{I}'_X,\;\mathcal{I}'_Y$.
  \end{algorithmic}
\end{algorithm}

After stacking $X$ and $Y$, we run PCA to extract the top $m$ principal directions. For each such $u^{(\ell)}$, we project points of $X$ and $Y$ onto $u^{(\ell)}$, sort the projected coordinates, and keep the indices corresponding to the smallest $k_X$ and largest $k_X$ projections in $X$, and similarly for $Y$. Altogether, this accumulates into $\mathcal{I}'_X$ and $\mathcal{I}'_Y$, which capture extremes along all principal axes.

\subsubsection{Main Procedure}
Finally, we combine the centroid-based indices and PCA-based indices, extract the union, and compute a fast ANN-based Hausdorff on the resulting subsets.

\begin{algorithm}[H]
  \footnotesize
  \caption{ProjHausdorff\(\bigl(A,B,\alpha\bigr)\)}
  \label{alg:main}
  \begin{algorithmic}[1]
    \REQUIRE $A\in\mathbb{R}^{n_A\times D},\;B\in\mathbb{R}^{n_B\times D},\;\alpha\in(0,1).$
    \ENSURE Approximation $\widehat H(A,B)$, and subset sizes $|A_{\mathrm{sel}}|,\;|B_{\mathrm{sel}}|$.
    \STATE Set $m \leftarrow \lfloor\sqrt{D}\rfloor,\;\alpha' \leftarrow \alpha/m.$
    \STATE $(\,\mathcal{I}_c^A,\mathcal{I}_c^B)\leftarrow\text{CentroidIndices}(A,B,\alpha)$.  
    \STATE $(\,\mathcal{I}_p^A,\mathcal{I}_p^B)\leftarrow\text{PCAProjIndices}(A,B,\alpha',m).$
    \STATE $\mathcal{I}^A \leftarrow \mathcal{I}_c^A \cup \mathcal{I}_p^A,\quad
            \mathcal{I}^B \leftarrow \mathcal{I}_c^B \cup \mathcal{I}_p^B.$
    \STATE Extract subsets $A_{\mathrm{sel}} = \{\,a_i: i\in\mathcal{I}^A\},\;B_{\mathrm{sel}}=\{\,b_j:j\in\mathcal{I}^B\}.$
    \STATE Compute 
      $\widehat h(A,B) = \max_{a\in A_{\mathrm{sel}}}\min_{b\in B_{\mathrm{sel}}}\|a-b\|$,
      $\widehat h(B,A) = \max_{b\in B_{\mathrm{sel}}}\min_{a\in A_{\mathrm{sel}}}\|b-a\|$,
      via a fast ANN (e.g., Faiss).
    \STATE $\widehat H(A,B)\leftarrow \max\{\widehat h(A,B),\,\widehat h(B,A)\}.$
    \RETURN $\widehat H(A,B),\;| \mathcal{I}^A|,\;| \mathcal{I}^B|.$
  \end{algorithmic}
\end{algorithm}

We first call \texttt{CentroidIndices} to keep extremes along the centroid direction. Then \texttt{PCAProjIndices} extracts extremes along the top $m$ PCA directions, using fraction $\alpha'= \alpha/m$ so that PCA extremes collectively do not exceed $\alpha\,(n_A + n_B)$ points. Taking unions yields two small subsets $A_{\mathrm{sel}}$ and $B_{\mathrm{sel}}$. Finally, we run a fast ANN-based Hausdorff on those subsets to obtain $\widehat H(A,B)$. By construction,
\begin{align*}
      &\widehat H(A,B)
  \;=\; 
  \max_{\ell=0,\dots,m}H_{u^{(\ell)}}(A,B),
  \quad\\
  &\ell=0\text{ corresponds to centroid direction},\;\\
  &\ell>0\text{ to PCA directions}.
\end{align*}

\subsection{Analysis}
\label{sec:analysis}

Let $n = n_A + n_B$, $m = \lfloor\sqrt{D}\rfloor$, and assume $\alpha \ll 1$ (e.g., $\alpha = 0.01$). For typical $D$ (a few to a few hundred) and $n$ (thousands to millions), we can also assume $D \ge \log n$. The algorithm proceeds in four phases: centroid projection, PCA projection, union–subset extraction, and ANN-based Hausdorff. Below we analyze each phase in turn, expressing costs in terms of $n$, $D$, $m$, and $\alpha$.

\paragraph{Centroid projection.} We compute centroids of $A$ and $B$ in $O(nD)$ time, form the direction vector $u\in\mathbb{R}^D$, and project all $n$ points onto $u$. Each projection is a dot-product costing $O(D)$, for a total of $O(nD)$. Sorting the $n$ scalar projections takes $O(n \log n)$, but since $D \ge \log n$, the overall cost is 
  \[
    O(nD + n\log n) \;=\; O(nD).
  \]

\paragraph{PCA projection.} We stack the $n$ points from $A$ and $B$ into an $n\times D$ matrix and compute its top $m$ principal components. A randomized or truncated SVD on an $n\times D$ matrix to extract rank $m$ costs 
  \[
    O\bigl(n\,D\,m\bigr)
    \;=\; 
    O\bigl(n\,D^{1.5}\bigr),
  \]
  since $m = \lfloor\sqrt{D}\rfloor$ and $D \ge 1$. Next, for each of the $m$ principal directions, we project $n$ points (cost $O(n)$ per direction, hence $O(mn)$ total) and sort the projections (cost $O(m\,n\log n)$). Because $D \ge \log n$, sorting $n$ numbers costs $O(nD)$. Therefore, projecting and sorting across all $m$ directions costs
  \[
    O\bigl(m\,n + m\,n\log n\bigr)
    \;=\; 
    O\bigl(m\,n\,D\bigr)
    \;=\; 
    O\bigl(n\,D^{1.5}\bigr).
  \]
  Altogether, the PCA projection phase is
  \[
    O\bigl(n\,D^{1.5} + n\,D^{1.5}\bigr)
    \;=\; 
    O\bigl(n\,D^{1.5}\bigr).
  \]

\paragraph{Union and subset extraction.} Let $\mathcal{I}_c^A,\mathcal{I}_c^B$ be the indices from centroid projection (each of size $O(\alpha\,n)$), and $\mathcal{I}_p^A,\mathcal{I}_p^B$ be the indices from PCA projection (each of size $O(\alpha\,m\,n)$). Forming the union 
  $\mathcal{I}^A = \mathcal{I}_c^A \cup \mathcal{I}_p^A$, 
  $\mathcal{I}^B = \mathcal{I}_c^B \cup \mathcal{I}_p^B$
  costs $O(n)$. Extracting the subsets 
  $A_{\mathrm{sel}} = \{\,a_i: i\in\mathcal{I}^A\}$,
  $B_{\mathrm{sel}} = \{\,b_j: j\in\mathcal{I}^B\}$
  involves copying $O(\alpha\,m\,n)$ points of dimension $D$, for a cost of 
  \[
    O\bigl(\alpha\,m\,n\,D\bigr)
    \;=\; 
    O\bigl(\alpha\,n\,D^{1.5}\bigr).
  \]

\paragraph{ANN-based Hausdorff on the subsets.} The selected subsets have size 
  $|A_{\mathrm{sel}}| + |B_{\mathrm{sel}}| 
    = O(\alpha\,m\,n) = O(\alpha\,n\,D^{0.5})$. Building an ANN index (e.g., Faiss FlatL2) on $O(\alpha\,n\,D^{0.5})$ points in $\mathbb{R}^D$ costs 
  \[
    O\bigl(\alpha\,n\,D^{0.5}\times D\bigr)
    \;=\; 
    O\bigl(\alpha\,n\,D^{1.5}\bigr).
  \]
  Querying that index to compute directed Hausdorff distances for each of the $O(\alpha\,n\,D^{0.5})$ points costs 
  \[
    O\bigl(\alpha\,n\,D^{0.5}\log(\alpha\,n\,D^{0.5})\bigr).
  \]
  Since $\log(\alpha\,n\,D^{0.5}) \le \log n \le D$, this query cost is bounded by 
  \[
    O\bigl(\alpha\,n\,D^{0.5}\times D\bigr)
    \;=\; 
    O\bigl(\alpha\,n\,D^{1.5}\bigr).
  \]
  Thus, the entire ANN phase also costs $O(\alpha\,n\,D^{1.5})$.

Combining all four phases, the total sequential runtime is
\[
  O\bigl(n\,D + n\,D^{1.5} + \alpha\,n\,D^{1.5} + \alpha\,n\,D^{1.5}\bigr)
  \;=\; 
  O\bigl(n\,D^{1.5}\bigr),
\]
because $D^{1.5} \ge D$ whenever $D \ge \log n$ and $\alpha \ll 1$.

\vspace{1ex}
\noindent\textbf{Space Complexity.} Storing the original clouds $A$ and $B$ requires $O(nD)$ memory. The projection arrays for centroid and PCA phases use $O(n)$. The selected subsets $A_{\mathrm{sel}}$, $B_{\mathrm{sel}}$ occupy $O(\alpha\,m\,n\,D) = O(\alpha\,n\,D^{1.5})$. The ANN index built on those subsets also requires $O(\alpha\,n\,D^{1.5})$. Since $\alpha \ll 1$, the dominant term is $O(nD)$.  

\vspace{1ex}
\noindent\textbf{Parallelism.} If $P$ CPU cores are available, each phase can be divided approximately evenly across $P$ threads: (i) the $n$ projections onto a single direction can be split so that each core processes about $n/P$ points, reducing $O(nD)$ to $O(nD/P)$; (ii) the PCA fitting and per‐direction projections/sorts can be parallelized across $P$ cores, reducing $O(nD^{1.5})$ to $O(nD^{1.5}/P)$; (iii) forming unions and extracting subsets of size $O(\alpha\,n\,D^{0.5})$ can be done in $O(n/P)$ and $O(\alpha\,n\,D^{0.5}/P)$ time, respectively; (iv) building and querying the ANN index on $O(\alpha\,n\,D^{0.5})$ points can be distributed, reducing $O(\alpha\,n\,D^{1.5})$ to $O(\alpha\,n\,D^{1.5}/P)$. Thus, on $P$ cores the total parallel runtime becomes $O\bigl(n\,D^{1.5}/P\bigr)$.

\vspace{1ex}
\noindent\textbf{Scalability.} Because the selected subset size scales as $O(\alpha\,m\,n)$, with $\alpha \ll 1$ and $m = \sqrt{D}$, the work on the small subset remains negligible compared to the $O(n\,D^{1.5})$ projection/PCA cost. As $n$ or $D$ grows, the method remains viable for moderately large $D\approx \log n$ (where $nD^{1.5}$ is subquadratic) and even for high $D$ when the data lie near a lower-dimensional subspace.  

\subsection{Theoretical Properties}
\label{sec:properties}

\subsubsection{Single-Direction Bound}
Fix a unit vector $u$. Define the 1D directed Hausdorff along $u$ by
\begin{align*}
  &h_u(A,B) 
  \;=\; 
  \max_{a\in A}\min_{b\in B}\bigl|\pi_u(a) - \pi_u(b)\bigr|,
  \quad\\
  &H_u(A,B) = \max\{h_u(A,B),\,h_u(B,A)\}.
\end{align*}
Then for any $u$,
\begin{align*}
      \label{eq:single_bound}
  &h_u(A,B)\;\le\;h(A,B)\;\le\;h_u(A,B)\;+\;2\,\delta(u),
  \quad\\
  &H_u(A,B)\;\le\;H(A,B)\;\le\;H_u(A,B)\;+\;2\,\delta(u),
\end{align*}
where $\delta(u)$ is defined in \eqref{eq:delta_def}.

\begin{proof}[Sketch of Proof]
Let $a^\star \in A$ be the point that realizes $h(A,B)$, and let $b^\star \in B$ be its nearest neighbor in $B$. Then 
\begin{align*}
      h(A,B)
  &= \|\,a^\star - b^\star\|\\
  &\le\;\bigl|\pi_u(a^\star)-\pi_u(b^\star)\bigr|
       + \|\Pi_{u^\perp}(a^\star)\|
       + \|\Pi_{u^\perp}(b^\star)\| \\
  &\le\; h_u(A,B)\;+\;2\,\delta(u).
\end{align*}
The lower bound $h_u(A,B)\le h(A,B)$ holds because 
$\bigl|\pi_u(a)-\pi_u(b)\bigr|\le\|a-b\|$ for every pair $(a,b)$. The same argument applies to $h(B,A)$.  
\end{proof}

\subsubsection{Multiple-Direction Bound}
Let $\mathcal{U} = \{\,u^{(0)},\dots,u^{(m)}\}$ be a finite set of unit directions. Define
\[
  H_{\mathcal{U}}(A,B) 
  = \max_{u\in\mathcal{U}}H_u(A,B).
\]
Then
\begin{equation}
  \label{eq:multi_bound}
  H_{\mathcal{U}}(A,B)\;\le\;H(A,B)\;\le\;H_{\mathcal{U}}(A,B)\;+\;2\,\min_{u\in\mathcal{U}}\delta(u).
\end{equation}

\begin{proof}[Sketch of Proof]
Since $H_u(A,B)\le H(A,B)$ for each $u$, the first inequality follows by taking the maximum over $u\in\mathcal{U}$. For the second, 
\begin{align*}
      H(A,B)\;&\le\;\min_{u\in\mathcal{U}}\bigl\{\,H_u(A,B)\;+\;2\,\delta(u)\bigr\} \\
        &\le\;\max_{u\in\mathcal{U}}H_u(A,B)\;+\;2\,\min_{u\in\mathcal{U}}\delta(u).
\end{align*}
\end{proof}

\subsubsection{Monotonicity and Convergence}
Suppose $\mathcal{U}_1 \subset \mathcal{U}_2$. Then 
\begin{align*}
      &H_{\mathcal{U}_1}(A,B) \;\le\; H_{\mathcal{U}_2}(A,B) \;\le\; H(A,B),
  \quad\\
  &0\;\le\; H(A,B)\;-\;H_{\mathcal{U}_2}(A,B)\;\le\;2\,\min_{u\in\mathcal{U}_2}\delta(u).
\end{align*}
As $\mathcal{U}_2$ grows (more directions), $\min_{u\in\mathcal{U}_2}\delta(u)$ can only decrease, so $H_{\mathcal{U}_2}(A,B)$ increases (approaching $H(A,B)$). In the limit that $\mathcal{U}$ becomes dense on the unit sphere, $\min_{u\in\mathcal{U}}\delta(u)\to 0$ and thus 
$\widehat H(A,B) = H_{\mathcal{U}}(A,B)\to H(A,B)$.

\subsubsection{High-Dimensional Robustness}
In very large $D$, random directions $u$ tend to yield $\delta(u)\approx$ (cloud radius), giving poor bounds. By contrast, the top principal component $u^{(1)}$ minimizes $\max_{p}\|\Pi_{u^\perp}(p)\|$. Empirically, choosing $\{u_{\mathrm{centroid}},\,u^{(1)},\dots,u^{(m)}\}$ yields $\min_{u}\delta(u)\ll$ (radius), so the additive gap $2\,\min_{u}\delta(u)$ is small.

\subsubsection{Underestimation Bias}
From \eqref{eq:multi_bound}, 
\[
  \widehat H(A,B) = H_{\mathcal{U}}(A,B)\;\le\;H(A,B).
\]
Hence our algorithm never overestimates the true Hausdorff distance. The gap $H(A,B) - \widehat H(A,B)\le 2\,\min_{u\in\mathcal{U}}\delta(u)$ quantifies the worst-case underestimation.

\section{Evaluation}
\label{sec:eval}

In this section, we present a comprehensive empirical study of our proposed \textbf{ProHD} method in comparison with exact and sampling-based baselines. We first describe the datasets, configurations, and baselines. We then report on overall effectiveness (error versus runtime) using the slower exact methods (EBHD, ZHD, and points-ruling-out), and explain why we omit the faster ANN-Exact in that subsection. In subsequent analyses, we use ANN-Exact as the representative exact method when evaluating scalability. Finally, we carry out parameter sensitivity tests and ablation studies to investigate how various parameters affect accuracy and efficiency.

\subsection{Datasets and Experimental Setup}

We evaluate on four categories of datasets: CIFAR-10\cite{CIFAR10Krizhevsky09learningmultiple} and MNIST\cite{MNISTdeng2012mnist} (after PCA), the Higgs\cite{HiggsBaldi:2014kfa} physics dataset, and randomly generated point clouds which are distributed uniformly in the unit cube $[0,1]^D$ with an offset of $0.1$. In all experiments, we treat Faiss-accelerated ANN (denoted “ANN-Exact”) as a fast exact method. Every approximate method (ProHD, Random Sampling, Systematic Random Sampling) also uses the same Faiss ANN routine on its selected subset, so that differences between approximate methods arise solely from the selection step.

Table~\ref{tab:datasets} summarizes the domain, embedding dimensions, and point-set sizes for each dataset.

\begin{table*}[t]
\centering
\caption{Dataset Characteristics}
\label{tab:datasets}
\begin{tabular}{|l|l|c|l|}
\hline
\textbf{Dataset}                & \textbf{Domain}   & \textbf{Dimension(s)}                  & \textbf{Size Configuration(s)}                                                                                                                                      \\ \hline
CIFAR-10 (PCA)                  & Image             & $\{2,4,8,16,32,64,128,256\}$           & 6\,000 points per class (10 classes $\Rightarrow$ 45 non-repeated pairs)                                                                                              \\ \hline
MNIST (PCA)                     & Image             & $\{2,4,8,16,32,64,128,256\}$           & 6\,000 points per class (10 classes $\Rightarrow$ 45 non-repeated pairs)                                                                                                        \\ \hline
Higgs Subsets                   & Scientific        & 28                                     & \begin{tabular}[c]{@{}l@{}}(100\,k,100\,k),\;(100\,k,50\,k),\;(100\,k,25\,k),\\
\;(100\,k,12.5\,k),\;(1\,M,1\,M)\end{tabular}                                                    \\ \hline
Random Clouds                  & Synthetic         & $\{2,4,8,16,32,64,128,256\}$           & \begin{tabular}[c]{@{}l@{}}(100\,k,100\,k) for all dims; \\ For $D=4$ additionally:\\
\;(100\,k,50\,k),\;(100\,k,25\,k),\;(100\,k,12.5\,k),\;(1\,M,1\,M)\end{tabular}                                                    \\ \hline
\end{tabular}
\end{table*}

\noindent\textbf{Shared Parameters.} In our ProHD method, only one internal parameter $\alpha\in(0,1)$ (the total fraction of points kept across centroid and PCA directions) must be chosen at runtime. We fix the number of PCA axes to $m=\lfloor\sqrt{D}\rfloor$, so specifying $\alpha$ fully determines the subset size. For fairness, both Random Sampling and Systematic Random Sampling select exactly $\lceil\alpha\,(n_A + n_B)\rceil$ points per set pair. We refer to $\alpha$ as the \emph{selection fraction}. In addition, the embedding dimension $D$ and the point-set sizes $(n_A,n_B)$ serve as global parameters that affect all baselines.

\noindent\textbf{Metrics.} All methods are evaluated according to two metrics:
\begin{itemize}
  \item \emph{Relative Error:} defined as 
  \[
    \text{Rel.\ Error} \;=\; \frac{\bigl|\widehat H - H\bigr|}{H}\times 100\%,
  \]
  where $H$ is the exact Hausdorff distance (via ANN-Exact) and $\widehat H$ is the approximation produced by each method.
  \item \emph{Runtime:} measured in seconds, including projection or sampling overhead, Faiss index construction, and Faiss querying. Whenever a method supports multithreading, we use all 64 CPU threads. If a baseline does not support multithreading, it runs single-threaded but still relies on the ANN-based method (Faiss) for distance computation.
\end{itemize}

\noindent\textbf{Baselines.} We compare against the following methods:
\emph{EBHD:}\cite{EBHD27053955} Exact Hausdorff via KD-tree for one nearest-neighbor query per point instead of nested scans. \emph{ZHD:}\cite{ZHD8125673} Zelinka’s optimized exact Hausdorff implementation. \emph{Points-Ruling-Out with Systematic Random Sampling (Ruleout):}\cite{RYU2021107857} An exact pruning-based algorithm that eliminates dominated points before computing distances. \emph{ANN-Exact:}\cite{ANNunknown} Faiss ANN on the full sets uses a FlatL2 index (brute force). Zero error, generally fast. \emph{Random Sampling:}\cite{RYU2021107857} Uniformly sample $\lceil\alpha(n_A+n_B)\rceil$ points from each set; compute Hausdorff on the sampled points via Faiss. \emph{Systematic Random Sampling:}\cite{RYU2021107857} Apply a random permutation to each set and take every $\lfloor1/\alpha\rfloor$-th point to reach roughly $\alpha(n_A+n_B)$ points; compute Hausdorff via Faiss. Notice that the first four baselines are exact while the last two are approximations. (We did not include the Greedy Net algorithm\cite{chubet2025approximatingdirectedhausdorffdistance} because the original work provides no implementation and our own attempt showed poor runtime and error performance. We include ZHD rather than OHD or NOHD because of similar pruning strategies, and because NOHD strictly requires non-overlapping points, which we cannot guarantee.)

We run exact baselines on all threads because they are very slow if run in sequential. We also let the approximate baselines to run only on a single thread as they can finish in a few seconds and parallelism could slow them down instead.

\subsubsection*{Environment} All experiments are implemented in Python and executed on CloudLab using machines running Ubuntu 22.04.2 LTS (GNU$/$Linux 5.15.0-131-generic x86\_64). Each node is equipped with two AMD EPYC 7543 32-core processors running at 2.80GHz, 256GB of ECC memory (16×16GB 3200MHz DDR4), a 480GB SATA SSD, and a 1.6TB NVMe SSD (PCIe v4.0). 

\subsection{Overall Effectiveness}
To illustrate the trade-off between speed and accuracy, we plot \emph{average relative error} versus \emph{runtime} for each method on CIFAR-10, MNIST, Higgs, and Random Clouds. In these plots, we include the slower exact methods—EBHD, ZHD, and Points-Ruling-Out—because they are significantly slower than ANN-Exact. (We omit ANN-Exact from the figures, noting only that it achieves zero error at runtimes below those of EBHD, ZHD, and Ruleout.) In subsequent sections, we adopt ANN-Exact as the representative exact method, since it is the fastest exact solution.

\begin{figure}[t]
    \centering
    \includegraphics[width=\linewidth]{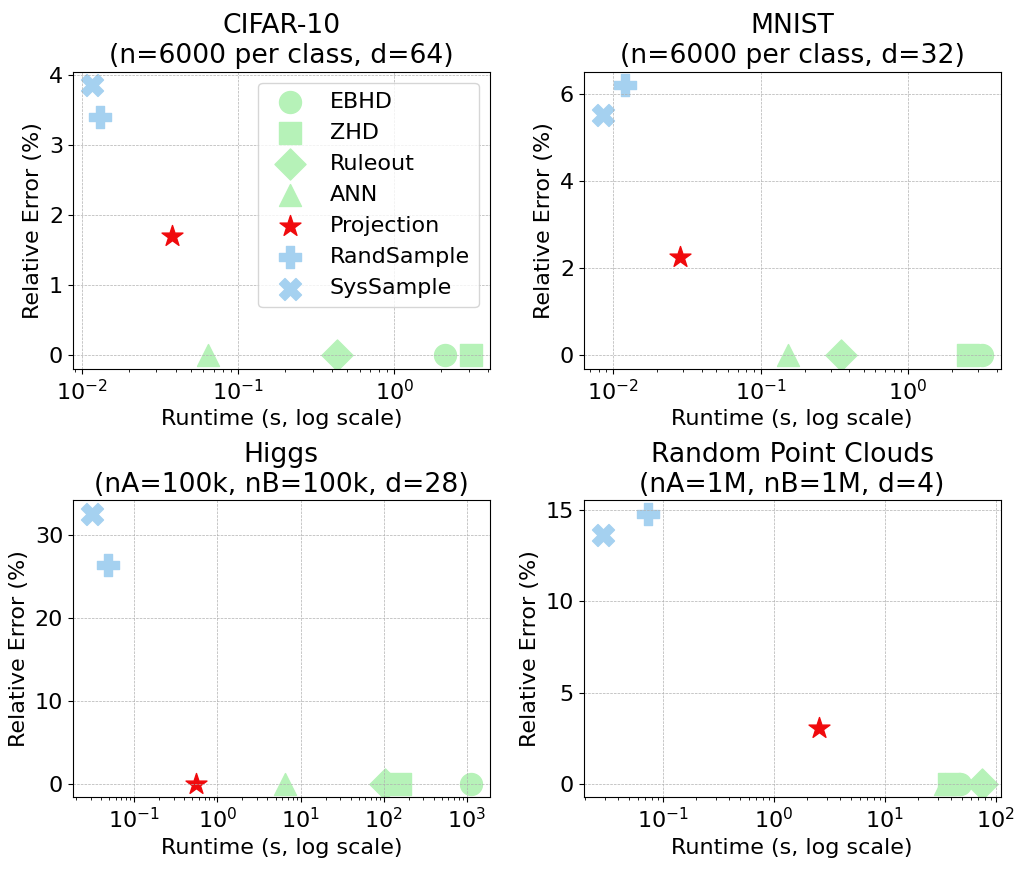}
    \caption{Average \emph{Relative Error} (\%) versus \emph{Runtime} (s, log scale) for EBHD, ZHD, Points-Ruling-Out (Ruleout), and three approximate methods (ProHD, Random Sampling, Systematic Sampling) on four datasets. We exclude ANN-Exact from these plots because it always achieves zero error at runtimes strictly lower than EBHD, ZHD, or Ruleout. }
    \label{fig:overall_effectiveness}
\end{figure}

\noindent\textbf{CIFAR-10 (PCA).} In the CIFAR-10 panel (top-left of Fig.~\ref{fig:overall_effectiveness}), both EBHD and ZHD achieve zero relative error but incur long runtimes—ranging from $1\,\mathrm{s}$ at $D=2$ up to over $60\,\mathrm{s}$ at $D=256$. Points-Ruling-Out (Ruleout) also has zero error but runs in $0.5\,\mathrm{s}$–$20\,\mathrm{s}$ depending on $D$. Among approximate methods, Random Sampling and Systematic Sampling complete in under $0.02\,\mathrm{s}$ but suffer high error (around $3.0\%$–$4.0\%$). Our ProHD method requires $0.02\,\mathrm{s}$–$0.20\,\mathrm{s}$ (increasing slightly with $D$) and achieves much lower error (between $0.05\%$ and $1.7\%$). In particular, at $D=64$, ProHD has $0.05\%$ error in $0.04\,\mathrm{s}$, whereas Random Sampling sees $3.3\%$ error in $0.02\,\mathrm{s}$.

\noindent\textbf{MNIST (PCA).} The MNIST panel (top-right) shows a similar trend: EBHD, ZHD, and Ruleout all attain zero error but take several seconds as $D$ increases. Random Sampling and Systematic Sampling remain under $0.02\,\mathrm{s}$ yet commit over $5\%$ error at $D\le 32$. ProHD consistently outperforms these sampling methods: for $D=32$, it incurs $0.2\%$ error in $0.06\,\mathrm{s}$, compared to $6.2\%$ (Random) and $6.1\%$ (Systematic) at $0.02\,\mathrm{s}$.

\noindent\textbf{Higgs ($D=28$).} In the Higgs panel (bottom-left), we plot five size configurations—(100k,100k), (100k,50k), (100k,25k), (100k,12.5k), and (1M,1M). EBHD, ZHD, and Ruleout all achieve zero error but range from $10\,\mathrm{s}$ at smaller sizes to over $300\,\mathrm{s}$ at 1M–1M. Random Sampling and Systematic Sampling finish in $0.02$–$3\,\mathrm{s}$ but have relative error above $24\%$ in every configuration. ProHD achieves $1.2\%$–$1.6\%$ error on 100k–100k through 100k–12.5k, and $2.3\%$ error at 1M–1M, with runtimes below $0.5\,\mathrm{s}$ in the 100k configurations and $4.8\,\mathrm{s}$ at 1M–1M.

\noindent\textbf{Random Clouds.} The Random Clouds panel (bottom-right) uses the same five size configurations. EBHD, ZHD, and Ruleout again yield zero error but take tens to hundreds of seconds. Random Sampling and Systematic Sampling run in under $0.10\,\mathrm{s}$ but incur $12\%$–$50\%$ error. ProHD’s error is $3.1\%$–$4.5\%$ at 100k sizes and $5.4\%$ at 1M–1M, with runtimes from $0.15\,\mathrm{s}$ to $1.8\,\mathrm{s}$.

\medskip
\noindent\textbf{Key Observations from Overall Effectiveness:}
ProHD consistently outperforms both Random Sampling and Systematic Sampling by large margins (often $>10\times$ lower error) while remaining within a small constant factor of their runtime. Exact methods EBHD, ZHD, and Ruleout achieve zero error but are up to two orders of magnitude slower than ProHD. Although we do not show ANN-Exact in Fig.~\ref{fig:overall_effectiveness}, it always attains zero error at a runtime that is strictly less than the slowest of EBHD, ZHD, or Ruleout; hence, for clarity, we omit it from these plots and use it as the canonical exact method in all subsequent scalability studies.

\noindent\textbf{Sample Efficiency Comparison.} 
\begin{table*}[t]
\centering
\caption{Sample Sizes Needed to Match ProHD’s Accuracy}
\label{tab:subset_comparison}
\begin{tabular}{|l|c|c|c|c|c|c|}
\hline
\textbf{Dataset (Configuration)} & \(\mathbf{D}\) & \textbf{ProHD Sample Size} & \textbf{Random Sample Size} & \(\frac{\text{Rand.}}{\text{ProHD}}\) & \textbf{Systematic Sample Size}   & \(\frac{\text{Sys.\ Rand.}}{\text{ProHD}}\)\\ \hline
MNIST (6k,6k)                    & 32             & 2\,450                      & 3\,990\,$(\pm 10)$               & 1.63                     & 8\,140\,$(\pm 10)$              & 3.33  \\ \hline
Higgs (100k,100k)                & 28             & 7\,182                      & 36\,000\,$(\pm 100)$             & 5.01                     & 43\,600\,$(\pm 100)$            & 6.07  \\ \hline
Random (100k,100k)               & 4              & 6\,000                      & 125\,052                           & 20.84                    & 158\,384                        & 26.40 \\ \hline
\end{tabular}
\end{table*}
To highlight how much more sample-efficient ProHD is compared to random sampling, Table~\ref{tab:subset_comparison} shows the number of points each method needs to match ProHD’s accuracy. For three representative scenarios, we list the subset size that ProHD selects and the larger subsets required by Random Sampling and Systematic Sampling to achieve the same relative error.

As the table indicates, ProHD selects a dramatically smaller subset while achieving a given accuracy level. For instance, on an MNIST pair, ProHD needs only 2\,450 points (20.4\% of the full set) to reach 0.5\% error, whereas Random Sampling requires roughly 4,000 points and Systematic Sampling over 8,000 points—over 3 times more than ProHD. On the Higgs 100k–100k sets, ProHD attains 1.2\% error with about 7,182 points (3.59\% of all points), but Random and Systematic Sampling need approximately 36k (18\%) and 43.6k (21.8\%) points, respectively. For a Random Cloud in 4-D with 100k points per set, ProHD’s subset is just 6,000 (3\%) for 3.1\% error, whereas Random Sampling uses 125k (62.5\%) and Systematic uses 158k (79.2\%) points for the same error. In summary, to match ProHD’s accuracy, competing methods must sample an order-of-magnitude more points, underscoring the significant gap in performance.

\subsection{Parameter Sensitivity Analysis}
\begin{figure*}[t]
    \centering
    \includegraphics[width=\linewidth]{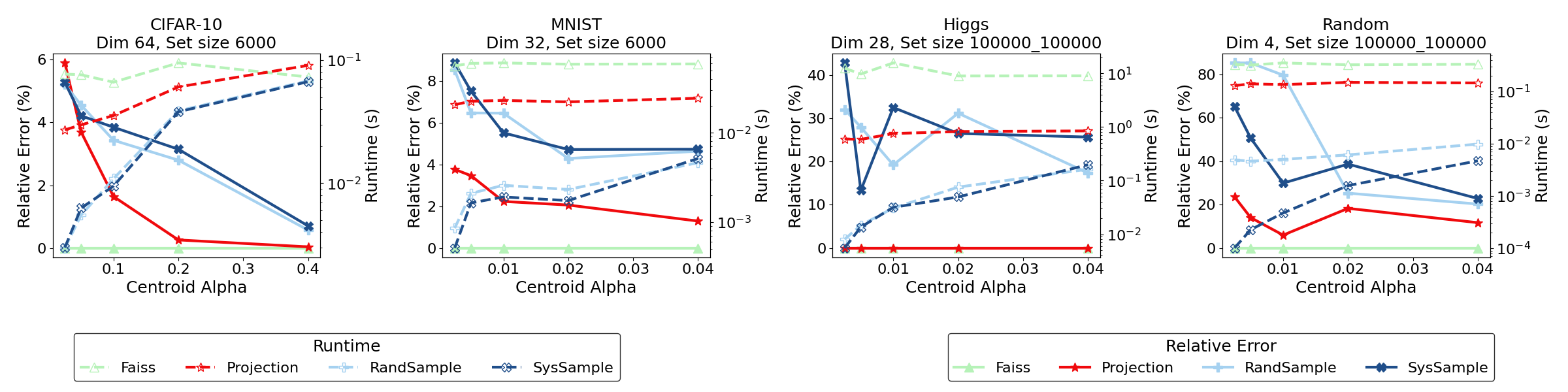}
    \caption{Parameter Sensitivity: \emph{Average Relative Error} (\%) (solid lines, left y-axis) and \emph{Runtime} (s) (dashed lines, right y-axis, log scale) versus selection fraction $\alpha$. Top row: CIFAR-10 ($D=64$, $n_A=n_B=6000$). Bottom row: Higgs ($D=28$, $n_A=n_B=100\,000$). ProHD’s error decreases sharply as $\alpha$ grows and remains much lower than Random and Systematic Sampling. Runtime of ProHD increases roughly linearly in $\alpha$, whereas sampling runtimes remain smaller until $\alpha$ becomes large.}
    \label{fig:param_sensitivity}
\end{figure*}
This analysis examines how the selection fraction $\alpha$ influences both approximation error and runtime. Figure~\ref{fig:param_sensitivity} shows average relative error and runtime versus $\alpha$ for two representative settings: CIFAR-10 ($D=64$, $n_A=n_B=6\,000$ per class pair) and Higgs ($D=28$, $n_A=n_B=100\,000$). All methods select exactly $\lceil\alpha(n_A+n_B)\rceil$ points from each set.

On CIFAR-10, ProHD’s relative error falls from $5.2\%$ at $\alpha=0.01$ to $1.6\%$ at $\alpha=0.05$, and drops below $0.5\%$ for $\alpha\ge0.08$. Random Sampling, by contrast, remains above $22\%$ error until $\alpha\ge0.10$. Systematic Sampling performs similarly to Random Sampling. ProHD’s runtime grows from $0.02\,\mathrm{s}$ at $\alpha=0.01$ to $0.10\,\mathrm{s}$ at $\alpha=0.20$, whereas Random and Systematic Sampling take $0.01\,\mathrm{s}$–$0.08\,\mathrm{s}$ over the same $\alpha$ range.  

On the Higgs dataset, ProHD’s error decreases from $24\%$ at $\alpha=0.005$ to $4.5\%$ at $\alpha=0.02$, and falls below $1\%$ at $\alpha=0.10$. Random Sampling’s error stays above $10\%$ until $\alpha\ge0.08$, and Systematic Sampling remains above $15\%$ even as $\alpha$ increases. ProHD’s runtime increases from $0.15\,\mathrm{s}$ at $\alpha=0.005$ to $0.60\,\mathrm{s}$ at $\alpha=0.10$, while sampling runtimes only rise from $0.04\,\mathrm{s}$ to $0.30\,\mathrm{s}$.  

These results confirm that ProHD achieves far lower error than uniform or systematic sampling for any given $\alpha$, and that its computational cost grows predictably with $\alpha$.

\subsection{Scalability Analysis}
We now investigate the scalability of our method along three dimensions: a) embedding dimensionality $D$, b) relative sizes of the two point sets, and c) the total number of points $n_A+n_B$. In all cases we fix $\alpha=0.01$ and compare ProHD to Random and Systematic Sampling, using ANN-Exact to compute ground truth.

\begin{figure}[t]
    \centering
    \includegraphics[width=\linewidth]{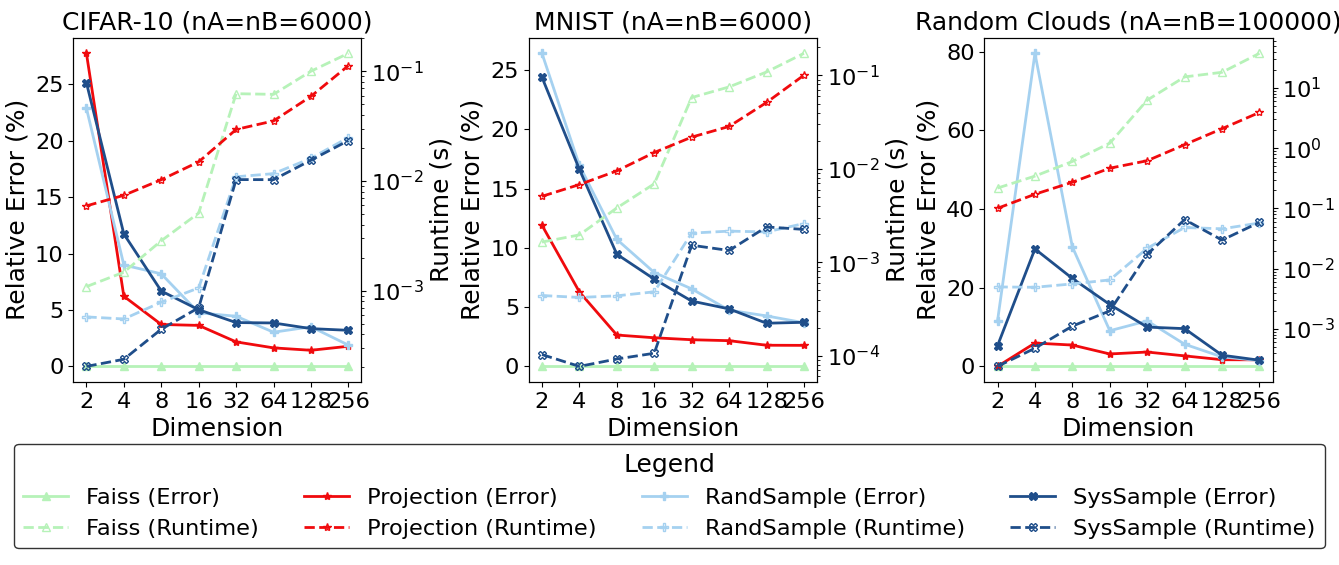}
    \caption{Dimension Scalability: \emph{Relative Error} (\%) (solid lines, left y-axis) and \emph{Runtime} (s) (dashed lines, right y-axis, log scale) versus embedding dimension $D$ for CIFAR-10 ($n_A=n_B=6000$), MNIST ($6000,6000$), and Random Clouds ($100\,000,100\,000$). ProHD’s error declines sharply with $D$, while runtime increases sublinearly. Random and Systematic Sampling maintain high error across all $D$.}
    \label{fig:ablation_dimension}
\end{figure}

\noindent\paragraph{Dimension Scalability.} Figure~\ref{fig:ablation_dimension} plots relative error and runtime as functions of $D$ for CIFAR-10 ($n_A=n_B=6\,000$), MNIST ($6\,000$ each), and Random Clouds ($100\,000$ each). ProHD’s error declines sharply as $D$ increases, while runtime grows sublinearly. For example, on CIFAR-10, ProHD’s error drops from $26.7\%$ at $D=2$ to $3.5\%$ at $D=8$, and is nearly $0\%$ by $D=64$, whereas Random and Systematic Sampling remain above $22\%$ error until $D=256$. On the Random Cloud, ProHD’s error falls from $45\%$ at $D=2$ to $2.8\%$ at $D=8$ and under $1\%$ by $D=64$, while the sampling methods still exceed $40\%$ error at $D=128$. Across datasets, ProHD’s runtime increases moderately (e.g., $0.005\,\mathrm{s}$ at $D=2$ to $0.12\,\mathrm{s}$ at $D=256$ on CIFAR-10), remaining practical even at the highest dimensions.

\begin{figure}[t]
    \centering
    \includegraphics[width=\linewidth]{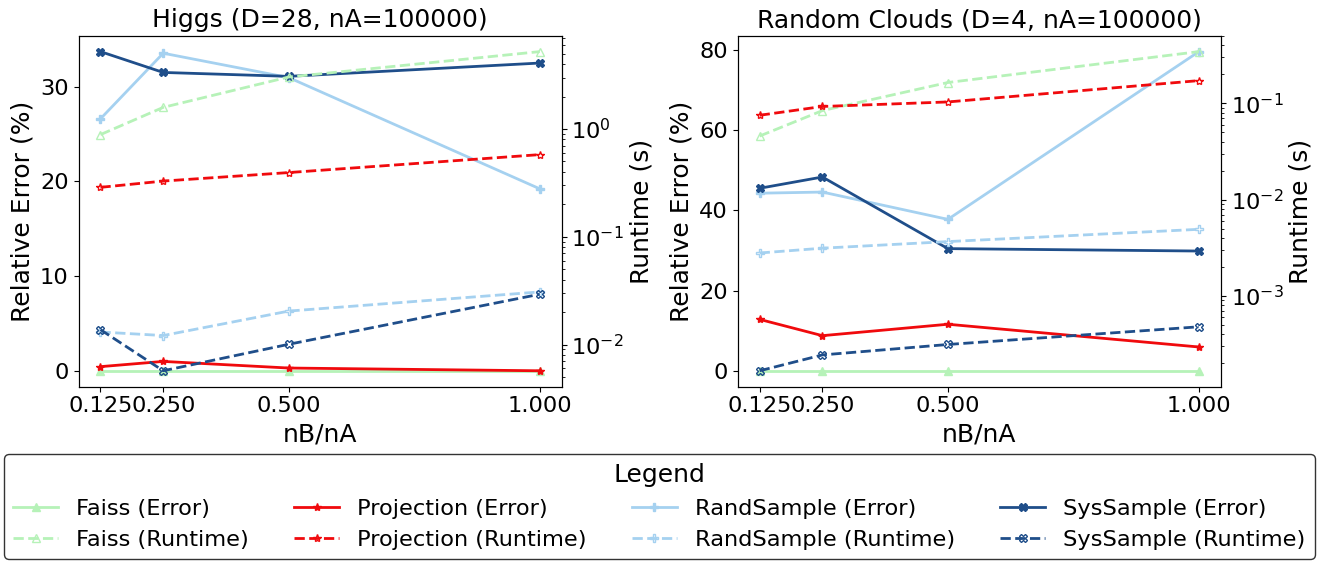}
    \caption{Set-Size Ratio Scalability: \emph{Relative Error} (\%) (solid lines, left y-axis) and \emph{Runtime} (s) (dashed lines, right y-axis, log scale) versus ratio $n_B/n_A$. Top: Higgs ($D=28$, $n_A=100\,000$). Bottom: Random Clouds ($D=4$, $n_A=100\,000$). ProHD maintains near-zero error on Higgs and substantially lower error on Random Clouds compared to sampling, even for imbalanced sets. ProHD’s runtime increases modestly with larger $n_B$.}
    \label{fig:ablation_set_size_ratio}
\end{figure}

\noindent\paragraph{Set-Size Ratio Scalability.} We next vary the ratio $n_B/n_A$ while keeping $n_A+n_B$ fixed. Figure~\ref{fig:ablation_set_size_ratio} shows results on the Higgs dataset ($D=28$, $n_A=100\,000$) and Random Clouds ($D=4$, $n_A=100\,000$) for $n_B/n_A \in \{0.125,0.25,0.5,1.0\}$. ProHD maintains near-zero error on Higgs and substantially lower error on Random Clouds compared to sampling, even for imbalanced sets. The sampling baselines incur large errors (e.g., $25\%$–$31\%$ on Higgs, $30\%$–$45\%$ on Random Clouds). ProHD’s runtime grows slightly as $n_B$ increases (from $0.12\,\mathrm{s}$ at $n_B/n_A=0.125$ to $0.24\,\mathrm{s}$ at parity on Higgs), while Random/Systematic remain around $0.05\,\mathrm{s}$.

\begin{figure}[t]
    \centering
    \includegraphics[width=\linewidth]{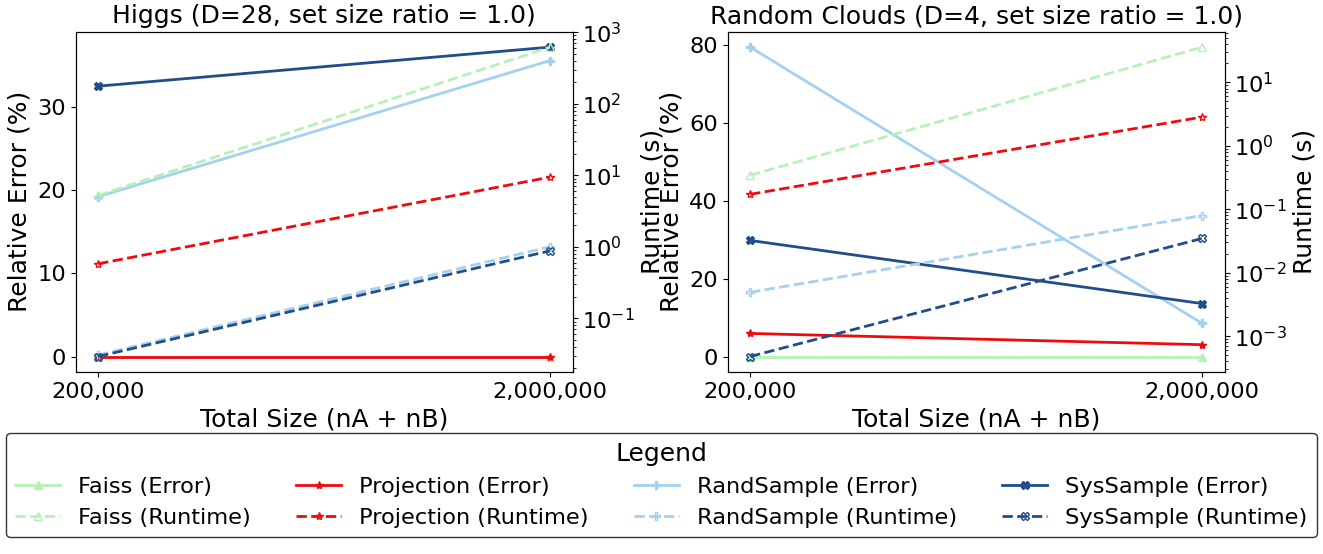}
    \caption{Total Set-Size Scalability: \emph{Relative Error} (\%) (solid lines, left y-axis) and \emph{Runtime} (s) (dashed lines, right y-axis, log scale) versus total size $n_A+n_B$. Left: Higgs ($D=28$). Right: Random Clouds ($D=4$). ProHD’s error stays low (below $0.5\%$ on Higgs, $6\%$ on Random Clouds) even at 2 million points, while runtime grows approximately linearly. Sampling methods show much higher error at both scales.}
    \label{fig:ablation_set_size}
\end{figure}

\noindent\paragraph{Total Set-Size Scalability.} Finally, we evaluate performance as the total number of points grows from $200k$ to $2M$. Figure~\ref{fig:ablation_set_size} reports results for Higgs ($D=28$) and Random Clouds ($D=4$) with $n_B/n_A=1.0$ (equal-sized sets). ProHD’s error stays low (below $0.5\%$ on Higgs, $6\%$ on Random Clouds) even at $2\times10^6$ points, far outpacing the $>25\%$ errors of sampling methods. Meanwhile, ProHD’s runtime increases roughly linearly with $n$, reaching $4.8\,\mathrm{s}$ at $2$ million points on Higgs and $1.8\,\mathrm{s}$ on Random Clouds, which is manageable given the dataset sizes. Random and Systematic Sampling are faster (under $0.3\,\mathrm{s}$ at $2$M) but provide much poorer accuracy.

Overall, these scalability experiments demonstrate that ProHD continues to deliver high accuracy at reasonable computational cost as dimensionality and dataset size increase, maintaining a wide accuracy gap over baseline approaches.

\section{Conclusion and Future Work}\label{sec:conclusion}
We presented \textbf{ProHD}, a fast, projection-based approximation to the Hausdorff distance that unifies centroid geometry, principal-component analysis, and modern ANN search. The method enjoys deterministic additive error bounds, runs in $O(nD^{1.5})$ sequential time, and requires only $O(nD)$ storage. Extensive experiments on image, physics, and synthetic data confirm that ProHD is up to two orders of magnitude faster than exact KD–tree or pruning algorithms while achieving $5$–$20\times$ lower error than uniform or systematic sampling at the same subset size.

\paragraph*{Future directions.} First, the additive bound in Eq.~\eqref{eq:final_bound} could be tightened by learning a compact \emph{direction dictionary} that further decreases $\min_{u}\delta(u)$ without increasing $|A_{\mathrm{sel}}|$. Second, GPU-resident Faiss indices or two-level coarse quantizers could shrink query latency, enabling real-time HD monitoring in high-throughput streams. Third, extending ProHD to other set distances—such as the directed Earth Mover’s or Gromov–Hausdorff—would broaden its applicability to non-Euclidean and metric-learning scenarios. Finally, we plan to study adaptive $\alpha$ schedules that accommodate strict error budgets while retaining constant-time execution.

\section{Acknowledgment}
This research is supported by the U.S.\@ Department of Energy (DOE) through
  the Office of Advanced Scientific Computing Research's ``Orchestration for Distributed \& Data-Intensive Scientific Exploration'' and
  the ``Decentralized data mesh for autonomous materials synthesis'' AT SCALE LDRD at Pacific Northwest National Laboratory.
PNNL 
is operated by Battelle for the DOE under Contract DE-AC05-76RL01830.
This work used TAMU ACES at Texas A\&M University through allocation CHE240191 from the Advanced Cyberinfrastructure Coordination Ecosystem: Services \& Support (ACCESS) program, which is supported by National Science Foundation grants \#2138259, \#2138286, \#2138307, \#2137603 and \#2138296.
Some results presented in this paper were partially obtained using the Chameleon testbed supported by the National Science Foundation.

\bibliographystyle{IEEEtran}

\end{document}